\documentclass[submission,copyright,creativecommons]{eptcs}
\providecommand{\event}{FESCA 2014} 
\usepackage{breakurl}               
\usepackage{acronym}
\usepackage{listings}
\usepackage{xcolor}
\usepackage{xspace}
\usepackage{acronym}
\usepackage{url}
\usepackage{verbatim}
\usepackage{multicol}
\usepackage{graphicx}

\title{Application of Ontologies in Identifying Requirements Patterns in Use Cases\thanks{This is partly funded by project LATiCES (ref. NORTE-07-0124-FEDER-000062), co-financed by the North Portugal Regional Operational Programme (ON.2 - O Novo Norte), under the National Strategic Reference Framework (NSRF), through the European Regional Development Fund (ERDF), and by national funds, through the Portuguese foundation for science and technology (FCT).}}
\author{Rui Couto
\institute{University of Minho\\ Braga, Portugal}
\email{ruicouto@di.uminho.pt}
\and
Ant\'onio Nestor Ribeiro
\institute{University of Minho\\ Braga, Portugal}
\email{anr@di.uminho.pt}
\and
Jos\'e Creissac Campos
\institute{University of Minho\\ Braga, Portugal}
\email{jose.campos@di.uminho.pt}
}
\def\titlerunning{Application of Ontologies in Identifying Requirements Patterns in Use Cases}
\def\authorrunning{R. Couto, A.N. Ribeiro \& J.C. Campos}
\begin{document}
\maketitle

\begin{abstract}
Use case specifications have successfully been used for requirements description. 
They allow  joining, in the same modeling space, the expectations of the stakeholders as well as the needs of the software engineer and analyst involved in the process.
While use cases are not meant to describe a system's implementation, by formalizing their description we are able to extract implementation relevant information from them.
More specifically, we are interested in identifying requirements patterns (common requirements with typical implementation solutions) in support for a requirements based software development approach.
In the paper we propose the transformation of Use Case descriptions expressed in a Controlled Natural Language into an ontology expressed in the Web Ontology Language (OWL).
OWL's query engines can then be used to identify requirements patterns expressed as queries over the ontology.
We describe a tool that we have developed to support the approach and provide an example of usage.
\end{abstract}

\definecolor{lightGray}{rgb}{0.93,0.93,0.93}

\lstset{ 
  backgroundcolor=\color{lightGray},
  basicstyle={\scriptsize \ttfamily},
  breakatwhitespace=false,
  breaklines=false,
  captionpos=b,
  commentstyle=\color{black},
  escapeinside={\%*}{*)},
  extendedchars=true,
  frame=single,
  keepspaces=true,
  keywordstyle=\color{blue},
  language=XML,
  morekeywords={*},
  numbers=left,
  numbersep=5pt,
  numberstyle=\tiny\color{gray},
  rulecolor=\color{black},
  showspaces=false,
  showstringspaces=false,
  showtabs=false,
  stepnumber=1,
  stringstyle=\color{black},
  tabsize=1,
  title=\lstname,
  belowskip=-1em,
  belowcaptionskip=1em,
}

\newcommand{\tab}{\hspace*{2em}}
\newcommand{\simpleFigure}[4]{ \begin{figure}[h] \centering \includegraphics[scale=#1]{#2} \caption{#3} \label{#4} \end{figure}}
\newcommand{\I}{{\tt <I>}\xspace}
\newcommand{\Ip}{{\tt <I>+}\xspace}
\newcommand{\R}{{\tt <R>}\xspace}
\newcommand{\T}{{\tt <T>}\xspace}
\newcommand{\D}{{\tt <D>}\xspace}
\newcommand{\etal}{{\it et al.}\xspace}
\newcommand{\protege}{Prot\'{e}g\'{e}\xspace}

\acrodef{ACE}{Attempto Controller English}
\acrodef{ANTLR}{ANother Tool for Language Recognition}

\acrodef{CNL}{Controlled Natural Language}
\acrodef{CNLs}{Controlled Natural Languages}
\acrodef{CLOnE}{Controlled Language for Ontology Editing}
\acrodef{CPN}{Coloured Petri Net}
\acrodef{CONON}{Context Ontology}
\acrodef{COBRA-ONT}{Context Broker Architecture}
\acrodef{CMS}{Content Management System}

\acrodef{DC}{Dublin Core}
\acrodef{DSL}{Domain Specific Language}

\acrodef{EUC}{Essential Use Cases}
\acrodef{WEF}{Workflow and Entities Framework}

\acrodef{RUS}{Restricted Use-Case Statements}

\acrodef{OMG}{Object Management Group}
\acrodef{OWL}{Web Ontology Language}
\acrodef{OWL-DL}{Web Ontology Language description logic subset }

\acrodef{RDF}{Resource Description Framework}
\acrodef{RDFS}{RDF Schema}

\acrodef{SWRL}{Semantic Web Rule Language}
\acrodef{SPARQL}{SPARQL Protocol and RDF Query Language}
\acrodef{SOUPA}{Standard Ontology for Ubiquitous and Pervasive Applications}

\acrodef{URI}{Uniform Resource Identifier}
\acrodef{UML}{Unified Modeling Language}
\acrodef{uCat}{Use Cases Analysis Tool}

\acrodef{W3C}{World Wide Web Consortium}

\acrodef{XSD}{XML Schema Definition}
\acrodef{XML}{Extensible Markup Language}

\section{Introduction}

	Developing software systems according to the stakeholders' expectations	is a well known problem in the software engineering area.
	One of  software engineering's major goals is to reflect the stakeholders concerns in the final solution, in order to provide useful software systems. 
	With traditional approaches \cite{Royce:1987:MDL:41765.41801}, requirements are specified at the beginning of the development process, and those specifications are used as guides for development.
	However, there is a gap between  user requirements and the software development process, which might result in the misunderstanding of the stakeholders' concerns. 

	Use cases are a popular method for requirements specification which contributes to solve this problem.
	They were proposed by Jacobson \cite{jacobson:uc} and later adopted by the \ac{OMG}.
	A use case model is composed of two parts: a graphical representation that summarizes the user interactions with the software system being described; and the specifications of each individual use case.
	
	Use cases mainly represent user functionalities and, as such, provide knowledge to derive high level information about the software systems they describe. 
	We aim to extract such information from the use cases, by means of an automated process.
	Hence, we present an approach to both formalize the use case specifications, and extract requirements patterns (which represent common concerns among several stakeholders) from them. 
	To enable this, use cases must be specified in a \ac{DSL}, which takes the form of a \ac{CNL}.
	In order to gain the reasoning leverage needed for identifying the patterns, we propose to use the inference capabilities of an ontology (in this case the \ac{OWL}). 
	To achieve this, we developed a transformation from use cases expressed in the adopted DSL into an OWL ontology.
	
	Our approach aims at several benefits on requirements based software development. 
	First, by providing a \ac{DSL}, it aims at allowing stakeholders to take part in the specification process. 
	The \ac{DSL} is at the same time formal and computable, yet, it is adequate for users to specify statements with due to being close to natural language. 
	Second, we aim also at improving the process of going from the requirements' specification to the final system. 
	The use of a formal DSL will allow us to predict errors and misunderstandings in requirements earlier in the software development process. 
	The assembling of architectural patterns \cite{Buschmann:1996:PSA:249013}, inferred from the requirements patterns, will support rapid prototyping of the system.
	At this stage, we have implemented a tool for requirements specification in order to explore these possibilities. 
	The tool supports the specification of user requirements in a textual format, using the defined \ac{DSL}. 
	It then transforms the requirements into an \ac{OWL} ontology, in which we are able to perform queries, and extract requirement patterns.
		
	In order to illustrate our approach, we present an example of an online \ac{CMS} for software models. We intent to equip it with functionalities requested by the stakeholders as use case descriptions.
	We will consider, for instance, the possibility to upload a model.
	
	The remaining of this work is organized as follows.
	Section~\ref{sect:rel_work} introduces  related work.
	In Section~\ref{sect:proposed_appr} we present the proposed approach with an example.
	We present our tool in Section~\ref{sect:tool}, again using the  example.
	In Section~\ref{sect:c_fw} we discuss the results achieved to date, and leave  suggestions for future work.
\section{Related work} 
\label{sect:rel_work}

\subsection{Languages for expressing use cases}

	Although there is no standard way to specify individual use cases, they are typically described in relatively free form textual formats (see, for example, Fowler \cite{Fowler:2003:UDB:861282} or Cockburn \cite{Cockburn:2000:WEU:517669}).
	One of the advantages of these textual descriptions of the scenarios composing a use case is that they allow for flexibility in the specifications.
	Unfortunately, they make it more difficult to formalize and operationalize the information in the scenarios.
	Furthermore, when  stakeholders take part in  use case specification, they typically use natural language to describe the scenarios.
	Such specifications can be ambiguous and lead the development team in error.

	Hence, approaches that propose analyzing use cases to extract implementation relevant information (e.g. architectural information) need to formalize languages in which use cases must be specified.
	Som\'{e} \cite{Some200643} presents an approach for software reengineering in which use cases play a central role as primary input, using a restricted form of natural language for use cases description.
	Similarly,  Biddle \etal describe \ac{EUC}~\cite{Biddle:2002:EUC:563801.563803}  as an approach  to operationalize use case descriptions.
	By reducing and restricting the language used in the use case descriptions, the authors were able to perform object oriented software development without intermediary transformations.
	The work by Biddle \etal is further supported with design patterns and composition techniques for \ac{EUC} \cite{Biddle:2001:patterns}.
	They present the specification for describing patterns for \ac{EUC}, as well as a pattern catalog.
	Executable use cases \cite{jorgensen:04:execuc} are another kind of restricted use cases, with computable capabilities, and with a defined mapping to \ac{CPN}.
	
	Our work takes from the above approaches the need to formally define a language for expressing use cases. 
	We propose an approach to specify the stakeholders' requirements in a \ac{CNL} \cite{Schwitter:2010:CNL:1944566.1944694}.
	The objective of using a \ac{CNL} is twofold.
	First, we want to allow stakeholders to take part in the requirements specification, providing a language understood by both  stakeholders and developers.  
	Second, we want to be able to transform use case specifications into other languages supporting the patterns identification mechanisms we want to achieve.
	We consider CNL to provide a good compromise between formalism and computability.

\subsection{Web Ontology Language (OWL)}

	In order to reason about the information present in the use case descriptions, we need some form of knowledge representation and inference mechanism.
	Ontologies	are a possibility to  achieve this.
	They provide the means to formally represent knowledge as a set of concepts, in a specific context.
	We are particularly interested in OWL \cite{owl:site}.	
	
	\ac{OWL} allows us to create domains for information specification,	and provides tools to reason about their knowledge.
	The \ac{W3C} specified three levels of formalism for the \ac{OWL} language. 
	The first level, \ac{OWL} Lite, is intended to simplify the specification of \ac{OWL} ontologies.
	The second level, \ac{OWL} DL, is intended to provide maximum expressiveness, while retaining the inference capabilities.
	Finally, the third level, \ac{OWL} Full, is a more complete language which preservers compatibility with \ac{RDF}, but has no reasoner available.
	Several syntaxes are available to write \ac{OWL} ontologies, as is the case of Manchester\footnote{Specification available at \url{http://www.w3.org/2007/OWL/wiki/ManchesterSyntax}, last accessed in 2013-12-11} and Sydney \cite{sydney2012:filipe}.

	All  \ac{OWL} ontology elements, and the ontology itself, contain an \ac{URI} which specifies the resource's location.
	Such web capability allows us to use and share domain concepts, among several locations.
	This not only promotes sharing and reuse of acquired knowledge among several projects, but also supports modelling web environments by providing all elements with a specific location.

The expressiveness of \ac{OWL} allow us to define requirements ontologies as well as their instances, and to perform queries over such knowledge. 
	While an ontology corresponds to a set of concepts for a specific domain, its instance is a set of concrete values, for those concepts.
	In \ac{OWL} those concepts are the classes (which are related between them), such as an \texttt{Actor}. 
	The concrete values for those classes are the individuals (c.f. instances of the classes), as for instance a \texttt{user}, which might be an instance of \texttt{Actor}.
	For \ac{OWL}, the \ac{W3C} specifies a set of components\footnote{Full list of allowed elements in \url{http://www.w3.org/TR/owl2-manchester-syntax/}, last accessed in 2013-12-11}, from which we highlight the following ones, due to their relevance for our purposes: 
	\begin{itemize}
		\item \texttt{Class} - A type (Class) of elements.
		\item \texttt{Individual} - A class instance.
		\item \texttt{ObjectProperty} - A relationship between two elements (for instance between two \texttt{Individual}).	
		\item \texttt{DataProperty} - Data associated with an \texttt{Individual}.
		\item \texttt{AnnotationProperty} - An annotation associated with an element.
		\item \texttt{DataType} - A specific kind of data (for instance \texttt{xsd:string}).
	\end{itemize}

	Several authors have proposed the use of OWL in support for software engineering. Dietrich and Elgar present an approach to formally describe design patterns in OWL \cite{1402019}. They perform a mapping between design patterns and OWL to enable design patterns' inference.
	The \ac{SWRL} rules provide the pattern inference capabilities.
	The work reported by Kirasi\'{c} and Basch is another example of pattern inference on \ac{OWL}, resorting to \ac{SWRL} rules \cite{Kirasic:2008:ODP:1430049.1430106}.
	 \ac{OWL} was also successfully used to extract other kind of knowledge, such as rationale informations from textual documents, as presented by L\'{o}pez \etal  \cite{Lopez:2012:BGS:2304796.2305263}.
	 Following these works, we intent to use \ac{OWL} to represent the knowledge present in use case descriptions, and then use its inference capabilities to identify patterns in their knowledge.

\subsection{From CNL to OWL}

	A number of approaches that map Controlled Natural Languages (CNLs) onto  ontologies, and onto OWL in particular,  can be found in the literature. 
	One example is the Rabbit language~\cite{Hart:2008:RDC:1789394.1789429}, defined to enhance communication between domain experts and ontology engineers, and the general purpose Sydney \ac{OWL} syntax \cite{sydney2012:filipe}.
	Other examples, are the Lite Natural Language, which is tailored for formal mapping and optimized for computational tasks \cite{bernardi:2007:lne}, and \ac{ACE}, a more expressive language than \ac{OWL} into which a mapping for a subset of \ac{OWL} has been defined \cite{kaljurand:phd:2007}.

	The work reported by Kuhn shows how controlled natural languages are easier to understand than the \ac{OWL} standard notations \cite{kuhn:2013:ace-owl}.
	His work is of special interest, since the author was able to successfully map an \ac{CNL} to \ac{OWL}, in order to improve OWL's readability \cite{kuhn:2013:ace-owl}.
	A list of \ac{CNL} statements is presented in \ac{ACE}. 
	Such statements have a respective \ac{OWL} mapping, which allows  mapping controlled english into declarative \ac{OWL} statements.
	This results in  statements which need less learning time and are more accepted by the users.

\section{The proposed approach}
\label{sect:proposed_appr}

\begin{figure*}[tbh]\small
		\centering
		\includegraphics[width=\textwidth]{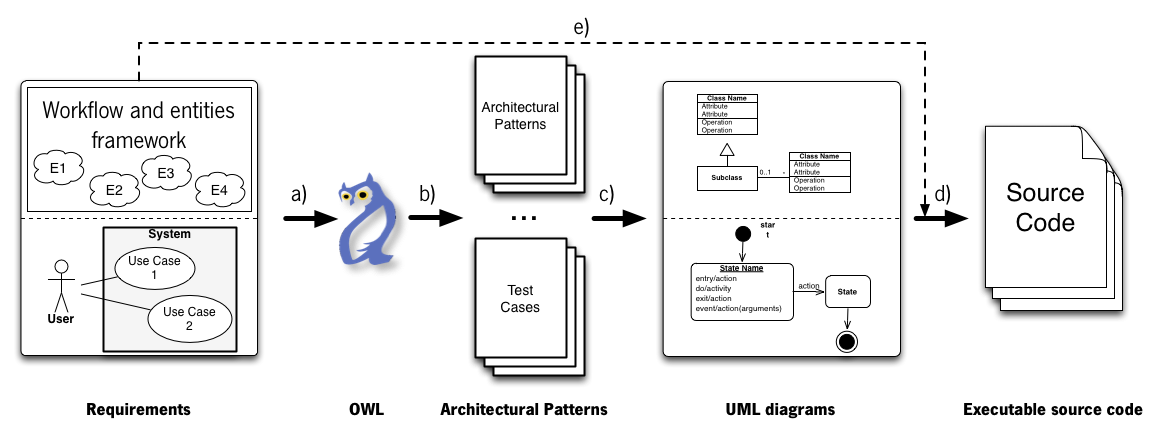}
		\caption{Overview of the proposed approach.}
		\label{figure:ov_process}
	\end{figure*}

	We propose a framework to extract common requirements' specifications from use case descriptions, as depicted in Figure~\ref{figure:ov_process}.
	In the figure, the full arrows represent the process flow, while the dashed arrow represents information flow.
	The use case allows the use of a restricted natural language, specifically, we propose the use of a \ac{CNL}. Without reducing flexibility, the \ac{CNL} improves the data extraction possibilities.

	Our process starts with the input of the use case descriptions (see Figure~\ref{figure:ov_process}, \textbf{Requirements}). 
In order to specify the use cases in a more controlled context,	we introduce the Workflow and Entities framework.
	Such framework indexes our tool to a specific domain. 
	When specifying the use cases, the authors resort to the entities in the domain, whose meaning is the same across all of the process.
	We consider \ac{OWL} to be adequate for our purposes for its declarative specifications and reasoning capabilities, compatible with our \ac{DSL} and our objectives.
	From the three \ac{OWL}'s levels of formalism, we are interested in the \ac{OWL-DL}, which has powerful inference capabilities and is supported by several reasoners.
	Figure~\ref{figure:ov_process} (phase \textbf{\ac{OWL}}) illustrates that we transform the use case specifications into \ac{OWL}.

	The expressiveness of  \ac{OWL} allow us to define requirements ontologies and to perform queries over such knowledge. 
	Resorting to the reasoners, we are able to query them, and extract patterns from them.
	We propose to extract requirement patterns from the use case specifications, after mapped onto \ac{OWL}.
	The transition \textbf{b)} and phase \textbf{Architectural Patterns} (Figure~\ref{figure:ov_process}) represent how we expect to extract the requirements patterns and integrate architectural patterns in the process.
 	We have developed a tool (described in Section~\ref{sect:tool}) which supports the ontology creation from use case descriptions, in order to validate our approach.

	The two last steps of the proposed approach, the \textbf{UML Diagrams} and the \textbf{Executable source code} steps, are future work.
	They represent the \ac{UML} diagrams (both structural and behavioral) to extract from the requirements patterns that, when assembled, will lead us to the source code.
	The dashed arrow \textbf{e)} represents how the entities and workflow framework provide the additional informations to achieve executable prototypes (as they contain the platform specific details).
	We detail these phases in Section~\ref{sect:c_fw}.
	
\subsection{The use case description}

	Typical use case descriptions are known to contain platform details, reflect user interface elements and, possibly, be ambiguous \cite{Biddle:2002:EUC:563801.563803}.
	In Figure~\ref{figure:use_case_login} we have an example of a use case specification (which should be provided by the stakeholders). 
	This is a traditional use case, where  the actor inputs and the system responses to such actions are specified. 
	Figure~\ref{figure:use_case_login} exemplifies how a user adds a new model to a web \ac{CMS}.
	We will refer to Figure~\ref{figure:use_case_login} during the remaining of this paper in order to present our approach.

	\begin{figure}[tbh]\small\center
		\begin{tabular}{|l|l|}
			\hline
			User (Actor) input & System (Actor) response\\
			\hline
			\hline
			Clicks in the New Model link & \\
			\hline
			 & Presents form with name, description, scope, language, file and image\\
			\hline
			Provides the requested fields & \\
			\hline
			Clicks the save button & \\
			\hline
			 & Validates the name, description, scope, language, file and image\\
			\hline
			 & Creates a new model \\
			\hline
			 & Lists all the models\\
			\hline
		\end{tabular}
		\caption{The model upload use case.}
		\label{figure:use_case_login}
	\end{figure}
		
	\subsubsection{Possible approaches}
		
	The verbosity of traditional use cases makes them hard to use in computational tasks.
	Due to the natural language complexity, it is possible to find all kind of details in the use cases, such as platform specific elements or elaborated text, which make the sentences harder to analyze.
	It is also possible to express the same thing in several distinct ways, and that makes it even harder to process the use case textual descriptions.
	The use case descriptions tend to provide other details than the required ones, which are unnecessary informations for our purposes.
	An alternative to  traditional textual use case specifications are  \ac{EUC} \cite{Constantine:1999:SUP:301248}. 
	They are simplified versions which aim to create more abstract and concise specifications. 
	In our approach, and as we are interested in creating computable representations, we aim to generate use case specifications similar to \ac{EUC}. 
	A way to achieve those specifications, is by guiding the user to create simple and objective statements in the use case description, mainly composed by the specific scenario elements.	

	To specify our simplified use cases descriptions, first we need a definition of a language. 
	An interesting language (\ac{ACE}) is presented by Fuchs and Schwitter \cite{fuchs99attemptoControlled}. 
	As discussed in Section~\ref{sect:rel_work}, Kuhn \cite{kuhn:2013:ace-owl} presents a study where the author describes how the \ac{ACE} language can be translated into \ac{OWL}, and more interestingly, how  \ac{ACE} improves the readability and understandability of the statements.
	The \ac{ACE} language targets at improving the specification of \ac{OWL} statements and knowledge scenarios description, and not at use case definition. 
	For that, we consider it unsuitable for our purposes, as we need a flexible language for use case descriptions.
	Inspired by  \ac{ACE}, we have defined a \ac{DSL} specification, which we have called the \ac{RUS} language.
	A \ac{RUS} file contains the valid statements for the use case description specifications. 
	It defines both the allowed input format, and the corresponding transformation of that specification to an ontology language.
	
	We consider that in our approach, the \ac{RUS} inherits from \ac{ACE} its two major strengths. 
	First it supports the writing of more natural and easier to understand statements. 
	Second it can be translated into \ac{OWL} statements.
	
	The \ac{ACE} language defines three kinds of tags.
	Individuals (the \I tag -- cf. \ac{OWL} entities), types (the \T tag -- cf. \ac{OWL} classes) and relationships (the \R tag -- cf. object properties).
	We added the \D tag for data properties, a type of element not covered in \ac{ACE}.	

	In Kuhns' work, a set of translations from \ac{ACE} directly into ``Manchester-Like Language'' (MLL)  is specified \cite{kuhn:2013:ace-owl}.
	While the Manchester syntax is a compact syntax to specify OWL ontologies, the MLL is a (non standard) restriction of that language.
	Based in those specifications, we specified the Manchester language equivalent to Kuhn's MLL.
	Since Kuhn has defined a transformation from \ac{ACE}, it is then possible to have transformations from \ac{ACE} to \ac{OWL} (Manchester language). 
	\ac{RUS} entries are descriptions of transformation rules from \ac{ACE} to Manchester language.
	Table \ref{table:ace_mll_owl} presents in the first column the \ac{ACE} format, in the second column the Manchester-Like Language representation and in the third column we defined the original Manchester Language statement.
	\begin{table}[tbh]\scriptsize\center
		\caption{Examples of statements in ACE, MLL and OWL.}
  		\begin{tabular}{|l|l||l|}
		  	\hline
			ACE & MLL & OWL Manchester\\
			\hline \hline
			\verb$<I1> <R> <I2>$ & \verb$<I1> <R> <I2>$ & \verb$Individual: <I1>$ \verb$ Facts: <R> <I2>$\\
			\hline
			\verb$<I1> does not <R> <I2>$ & \verb$<I1> not <R> <I1>$ & \verb$Individual:$ \verb$ <I1> Facts:$ \verb$   not <R> <I2>$\\
			\hline
			\verb$<I> is a <T>$ & \verb$<I> hasType <T>$ & \verb$Individual: <I>$ \verb$ Types: <T>$\\
			\hline
		\end{tabular}
		\label{table:ace_mll_owl}
	\end{table}

\subsubsection{The RUS language}

	The \ac{RUS} is a flexible template language, which allows developers to specify formats for use case descriptions.
	Each \ac{RUS} is composed by two parts, the input format and the target format.
	 
	The first part (input format) has two components: the placeholders and the keywords. 
	The placeholders are words (between ``$<$'' and ``$>$'') which mark the locations where the user will input text. 
	That text will result in data for the ontology.
	If we want a placeholder to accept multiple values, it should have the $+$ sign on front of it.
	The keywords are free text words which help to improve the statements readability. 
	This input format can have two or three placeholders depending on the users' needs, and as many keywords as required.
	This first part defines the format in which the user must write the statements. 
	
	The second part (the target format) is a set of four comma separated components (4-tuple), corresponding to the final format of the user's input.
	The first component is the ontology entity which the use case will affect. 
	The second component is the placeholder corresponding to to the actual entity (from the first part of the \ac{RUS} statement).
	The third component is the property that will be attributed to the second component, and the fourth component is its value.

	Briefly, the first component identifies how the user should write the requirements in \ac{CNL}. 
	It enables the validation of the use case description input, and at the same time the generation of parsing rules.
	The second component allows the specification of  how the extracted informations (specified according to the first component) should be mapped into the final ontology.

	Listing~\ref{figure:lsf_file} presents an example of a \ac{RUS} file that allows the specification of the model upload use case.
	From the \ac{RUS} file, we are able to extract two kinds of information.
	First, it is possible to extract a grammar, which produces a parser able to extract individuals, relationships, types and properties from the user specifications.
	Second, we are able to extract a set of regular expressions, which allow the runtime validation of the user requirements' statements.
\vspace{0.5em}
\begin{lstlisting}[caption=Example of RUS file., label=figure:lsf_file]
<I> <R> <I> -> Individual:,<I>,Facts:,<R> <I>          //action on individual
<I> <R> : <I>+ -> Individual:,<I>,Facts:,<R> <I>+      //action on several individuals
<I> _has <D>->Individual:,<I>,Facts:,<D>""^^xsd:string //individual has a property
\end{lstlisting}\vspace{-1.5em}

\subsubsection{RUS example}

	Consider the entries in Listing~\ref{figure:lsf_file}.
	The elements between ``$<$'' and ``$>$'' are the placeholders (where the user should input text). Words proceeded by an underscore (for instance \texttt{\_has} in line 3) are mandatory keywords.
	The arrow (\texttt{->}) separates the two components. 

	Regarding the entry in line 1, the first component states that an individual (\I) is related (\R) to another individual (second \I). 
	The second component, which specifies the target format, is specified with a 4-tuple of elements, separated by commas: 
	\texttt{Target\_Entity\_Description}, \texttt{Target\_Entity}, \texttt{Entity\_Pro\-perty} and \texttt{Property\_Value}.
		In this 4-tuple, the element \texttt{Target\_Entity\_Description}  specifies which \ac{OWL} element is this information related to (in the example, an \texttt{Individual}).

	The \texttt{Target\_Entity} element is the instance element, of the type \texttt{Target\_Entity\_Description}. 
	The element must be a placeholder (\I, \R, \D or \T, from the first component of the \ac{RUS} instruction).
	In this example, it is an instance of an \ac{OWL} individual (\I).

	The \texttt{Entity\_Property} element corresponds to a property of the \texttt{Target\_Entity}. 
	For an \texttt{Individual}, it is possible to have properties such as \texttt{Types} and \texttt{Facts}. In the example a \texttt{Fact} is being described. 
	It specifies where the \texttt{Property\_Value} element should be added at the target element.

	The last element, is the \texttt{Property\_Value}, which is an \texttt{Entity\_property} value. 
	It should contain the placeholders, and may also contain other \ac{OWL} allowed keywords (for instance \texttt{not} and \texttt{or}). 

	The presented entry allows us to declare statements identifying that an individual performs an action on another individual. 
	For example: \texttt{user clicks newModel}. 
	According to the rule, this entry is translated into: \texttt{Individual:,user,Facts:,clicks newModel}.

	This example states that an \texttt{user} (first individual) \texttt{clicks} (relationship) the \texttt{newModel} (second individual).
	As specified by the entry, this statement is translated into the following Manchester OWL statement:
\newpage
{\small
\begin{lstlisting}
Individual: <http://url/Req#user>
	Facts: <http://url/Req#clicks> <http://url/Req#newModel>
\end{lstlisting}
}\vspace{-1em}

	It is also possible to specify Restricted Use Case (RUS) statements with a variable number of entries by appending '\texttt{+}' to the statement, as illustrated in the second line of Listing~\ref{figure:lsf_file}.
	 An example of a (simplified) statement which relates an \texttt{Individual} (\I) with a set of \texttt{Individuals} (\Ip) is illustrated by:
\vspace{-0.5em}
\begin{lstlisting}
user inserts : name, description
\end{lstlisting}\vspace{-1em}

	This input states that the \texttt{user} (\I) \texttt{provides} (\R) the \texttt{name} and the \texttt{description} (\Ip).
	This input is divided into several equivalent statements, as presented next.
\vspace{0.5em}
\begin{lstlisting}
user inserts name
user inserts description
\end{lstlisting}\vspace{-1em}

	These statements can now be easily mapped into \ac{OWL} as presented in Listing~\ref{input:eq_owl} (see Section \ref{sect:ont_inst}), as they can be translated to the 4-tuple format (with the entry on the first line, in Listing~\ref{figure:lsf_file}).
	
\vspace{0.5em}
\begin{lstlisting}[caption=OWL example of an individual's facts., label=input:eq_owl]
Individual: <http://url/Req#user>
	Facts:
		<http://url/Req#inserts> <http://url/Req#name>,
		<http://url/Req#inserts> <http://url/Req#description>
\end{lstlisting}
	\begin{figure}[htbp]
		\center
		\scriptsize
		\begin{tabular}{|l|l|}
		\hline
		User input & System response\\
		\hline
		\hline
		user clicks newModel & \\
		\hline
		 & system requests : name, description, scope, language, file, image\\
		\hline
		user inserts : name, description, scope, language, file, image &\\
		\hline
		user clicks save & \\
		\hline
		 & system validates : name, description, scope, language, file, image\\
		\hline
		 & system creates model\\
		\hline
		 & system list models\\
		\hline
		\end{tabular}
		\caption{Formatted use case.}
		\label{figure:processed_usecase}
	\end{figure}
	\vspace{-0.5em}
	Rewriting the use case presented in Figure~\ref{figure:use_case_login} according to the \ac{RUS} in Listing~\ref{figure:lsf_file}, we obtain the description in Figure~\ref{figure:processed_usecase}.
	This formatted use case description corresponds to the input of our framework, represented as the phase \textbf{Requirements}, in Figure~\ref{figure:ov_process}.
\vspace{-1em}
\subsection{Extraction of entities}
\vspace{-0.5em}
	Given a use case as the one in Figure~\ref{figure:processed_usecase}, in the first steps of our approach we are interested in extracting the entities and 4-tuples from the use case description.
	There are two distinct inputs in this step. The first one, is the specification of the allowed language, the \ac{RUS}. 
	The second one is the use case itself.
	The \ac{RUS} allow us to create the grammar, which will extract informations from the use case.
	From the grammar file, a set of parsers able to extract the entities and 4-tuples from the requirements are achieved.
	Currently, it is possible to create two kinds of parsers.
	The first one allows the extraction of the list of entities, relationships, individuals and data properties, contained in the requirements specification.
	The second one, allows the extraction of the set of 4-tuple statements.
		
	With the \ac{RUS} in Listing~\ref{figure:lsf_file}, we were able to extract from the use case in Figure~\ref{figure:use_case_login}, the set of entities presented in Listing~\ref{figure:lst_entitites}, where \texttt{r} represents relationships (Object Properties) and \texttt{i} stands for Individuals.
	\vspace{-0.5em}
\begin{lstlisting}[caption=Extracted entities., label=figure:lst_entitites]
r:clicks,requests,inserts,validates,creates,list,
i:user,newModel,system,name,description,scope,language,file,image,save,model,models,
\end{lstlisting}
\vspace{0.5em}
	We were also able to extract the 4-tuples presented below:\vspace{-0.3em}
	\vspace{0.5em}
\begin{lstlisting}
Individual:,user,Facts:,clicks newModel
Individual:,system,Facts:,requests name
Individual:,system,Facts:,requests description
Individual:,system,Facts:,requests scope
Individual:,system,Facts:,requests language
Individual:,system,Facts:,requests file
Individual:,system,Facts:,requests image
Individual:,user,Facts:,inserts name
Individual:,user,Facts:,inserts description
Individual:,user,Facts:,inserts scope
Individual:,user,Facts:,inserts language
Individual:,user,Facts:,inserts file
Individual:,user,Facts:,inserts image
Individual:,user,Facts:,clicks save
Individual:,system,Facts:,validates name
Individual:,system,Facts:,validates description
Individual:,system,Facts:,validates scope
Individual:,system,Facts:,validates language
Individual:,system,Facts:,validates file
Individual:,system,Facts:,validates image
Individual:,system,Facts:,creates model
Individual:,system,Facts:,list models
\end{lstlisting}
\vspace{-1em}
	With this information, it is now possible to create an \ac{OWL} ontology with the previously presented methodology. 
	However, in order to create a complete instance,  some more informations  are required, related with the entities' types.
\vspace{-1em}
\subsection{Individuals and classes}
	Unless explicitly specified in the user requirements, both the classes and the individuals' type information are missing. 
	The individual to class association is crucial to create an ontology instance.
	In order to create it, the user must then specify all the relevant classes, and associate them with the individuals.
	
	In our example, we propose a set of classes derived from the problem domain, as presented next.

\vspace{0.5em}
\begin{lstlisting}
Actor, Link, Field, Text, Object,
\end{lstlisting}\vspace{-1em}

	The rationale for defining theses classes becomes clear when we specify the individuals to classes associations, as presented in Listing \ref{figure:individuals_types}.
	With these informations, it is now possible to create an ontology and an instance.

\vspace{0.5em}
\begin{lstlisting}[caption=Individual to classes association., label=figure:individuals_types]
user         (is an) Actor                 |     language     (is a)  Field (and a) Text
newModel     (is a)  Link                  |     file         (is a)  Field
system       (is an) Actor                 |     image        (is a)  Field
name         (is a)  Field (and a) Text    |     save         (is a)  Link
description  (is a)  Field (and a) Text    |     model        (is an) Object
scope        (is a)  Field (and a) Text    |     models       (is an) Object
\end{lstlisting}

\subsection{Ontology and instance}
	\label{sect:ont_inst}
	
	In order to create the ontology, we have created a module which resorts to the Apache Jena framework\footnote{Available at http://jena.apache.org/, last accessed in 2013-12-04}. 
	This framework allows us to take the previously extracted information, map it into an \ac{OWL} metamodel and create the source code.	
		
	Usually, the first element of an \ac{OWL} ontology is a set of prefixes, which represent additional resources to enhance the ontology.
	Example of prefixes are the \ac{XSD}, \ac{OWL}, \ac{XML}, \ac{RDF}, \ac{DC} and \ac{RDFS}. 	
	For our purposes, some of these prefixes, such as \ac{OWL} are mandatory, since we are defining \ac{OWL} ontologies.	
	Also, we need the \ac{RDF} prefix, which is used as a flexible data representation for domain resources. 
	Finally, after the prefixes, we can specify the ontology components. 
	
	With these informations, we were able to implement the tool which allows the creation of the ontology and respective instance (phase \ac{OWL} in Figure~\ref{figure:ov_process}).
	Resorting to the \ac{OWL} standards, we are able to create ontologies which can be integrated and processed in other tools, such as the \protege\footnote{http://protege.stanford.edu, last accessed in 2013-12-09} ontology editor.
	
	In order to exemplify our approach, we will resort to the first line of the use case description in Figure~\ref{figure:processed_usecase}:
\vspace{0.5em}
\begin{lstlisting}
user clicks newModel
\end{lstlisting}\vspace{-1em}	

	The described entry matches the rule in line 2 in the \ac{RUS} file presented in Listing~\ref{figure:lsf_file}.
	From this rule it is possible to know which elements we should extract, namely individuals (\texttt{user} and \texttt{newModel}) and relationships (\texttt{clicks}). 
	Next, the user should provide the types for the individuals. 
	In this case, we have two individuals, the \texttt{user} and the \texttt{newModel}. 
	Their types are \texttt{Actor} and \texttt{Link}, respectively (see Listing~\ref{figure:individuals_types}).
	With that information, it is finally possible to extract the following 4-tuple, which can be directly mapped into OWL.
	\vspace{0.5em}
\begin{lstlisting}
Individual:,user,Facts:,clicks newModel
\end{lstlisting}\vspace{-1em}

	The two first elements of the 4-tuple (\texttt{Individual:,user}), inform that we have an individual, named \texttt{user}. 
	This information is reflected in the following \ac{OWL} code.
\vspace{0.5em}
\begin{lstlisting}
Individual: <http://url/Req#user>
\end{lstlisting}\vspace{-1em}

	The three last elements in the 4-tuple (\texttt{user,Facts:,clicks newModel}) state that the user, on the property \texttt{Facts}, will have the pair value \texttt{'clicks newModel'}.
\vspace{0.5em}
\begin{lstlisting}
Individual: <http://url/Req#user>
  Facts: <http://url/Req#clicks> <http://url/Req#newModel>
\end{lstlisting}\vspace{-1em}	

	Finally, we have stated before that the user has the type \texttt{Actor}, hence we are able to reflect such information in \ac{OWL} as follows.
\vspace{0.5em}
\begin{lstlisting}
Individual: <http://url/Req#user>
  Types: <http://url/Req#Actor>
  Facts: <http://url/Req#clicks> <http://url/Req#newModel>
\end{lstlisting}\vspace{-1em}

	In Figure~\ref{figure:res_ontology} we have an extract of the resulting ontology. 
	The circles represent the classes, the diamonds individuals.
	The arrows between two classes are the \texttt{SubClass} (inheritance) relationship and the arrows between classes and individuals are the \ac{OWL} class association (for instance, an \texttt{user} \textit{has type} \texttt{Actor}). 
	The dashed arrow between the individuals represents the \texttt{clicks} relationships. 

\simpleFigure{0.37}{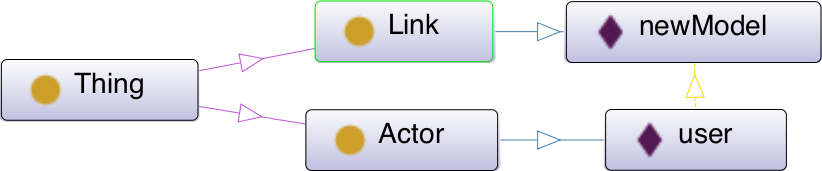}{Extract of the resulting ontology.}{figure:res_ontology}

\subsection{Pattern inference}

	\ac{OWL} provides several query languages, such as the \ac{SWRL} and the \ac{SPARQL}.
	In our approach we use \ac{SPARQL} to query the use case specification data, mapped onto \ac{OWL}.
	In order to infer the patterns, we describe the requirements patterns as \ac{SPARQL} queries.
	Then, we apply those queries in the use case descriptions.

	We present in Listing~\ref{figure:rp_login} a query  which identifies the requirement to have a model upload functionality.
	The \ac{SPARQL} language allow us to perform two kind of queries. 
	On the one hand, the \texttt{Select} query, allow us to ask questions about an ontology and get results.
	On the other hand, there is the \texttt{Ask} query, which verifies if a statement is valid for a given ontology.
	
	We query the knowledge base with the \texttt{Ask} statement, in order to identify if there is a user (\texttt{?actor}) which clicks in a link (\texttt{?link}) at a given web application (line 4), and that process results in a new model (\texttt{?result}, on line 5).
	The \texttt{?user} will identify which \texttt{Individual} \texttt{clicks} on a given link.
	The \texttt{?system} will relate the system's requested data with the \texttt{Field} type.
	Also, the \texttt{?system} will create the \texttt{?result}, which should be a new \texttt{model}.
	This pattern requests also that the user does not exit the current page (i.e., perform the exit action), represented with the \texttt{FILTER NOT EXISTS} statement (on line 6).
\vspace{0.5em}
\begin{lstlisting}[caption=Model upload requirement pattern., label=figure:rp_login]	
PREFIX ont: <http://www.url.com/Requirements/>
PREFIX rdf: <http://www.w3.org/1999/02/22-rdf-syntax-ns#>
ASK {
  ?actor ont:clicks ?link   . ?system ont:requests ?field .
  ?field rdf:type ont:Field . ?system ont:creates ?result . 
  FILTER NOT EXISTS {?user ont:exit ?link}
}
\end{lstlisting}

	The requirements pattern described in Listing~\ref{figure:rp_login}, applied to the use case in Figure~\ref{figure:processed_usecase}, returns the \texttt{true} value. 
	The \texttt{true} value means that the rule is satisfied in the given ontology, therefore we have the model upload pattern in the user requirements.
	
	The requirements patterns themselves provide informations about the user required functionalities. 
	Such functionalities provide hints about implementation details. 
	With more elaborated patterns and specifications, we expect to be able to associate architectural patterns with those requirements patterns. 
	We are still exploring the \ac{OWL} inference capabilities in order to mature our approach and to identify other data extraction possibilities.

\section{The Use Case Analysis Tool}
	\label{sect:tool}
	
	As already stated, the proposed framework is supported by a tool which guides the users when using our approach.
	We resort to the use case specified in Figure~\ref{figure:processed_usecase} in order to demonstrate our tool.
	We use the \ac{RUS} in Listing~\ref{figure:lsf_file} to specify the input language.	
	
	\begin{figure}[h!]\small
		\centering
		\includegraphics[width=50em]{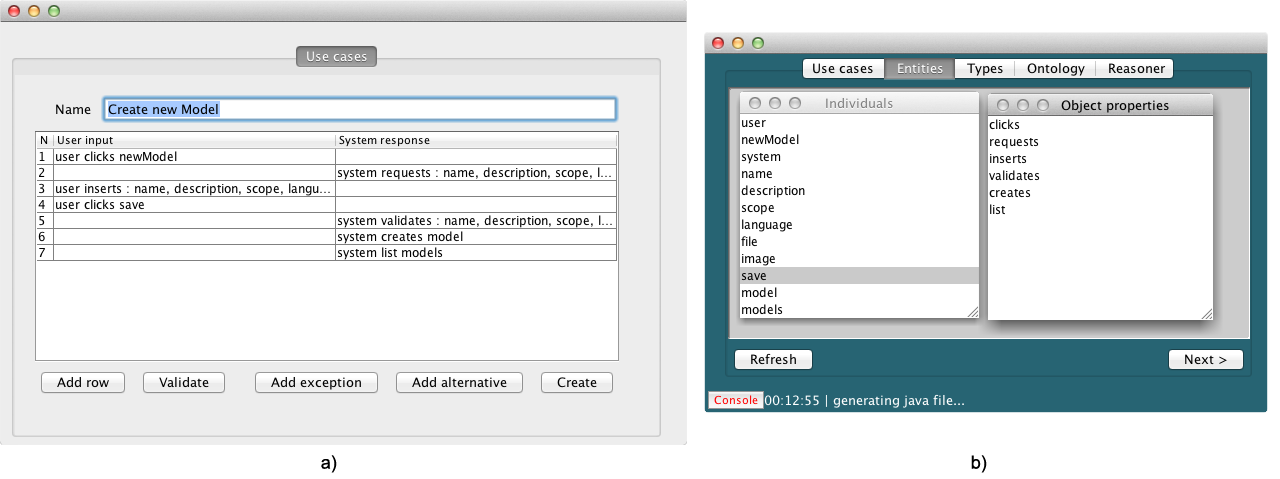}
		\caption{Use case description and inferred entities interfaces.}
		\label{figure:ss1}
	\end{figure}
	
	\begin{figure}[tb]\small
		\centering
		\includegraphics[width=30em]{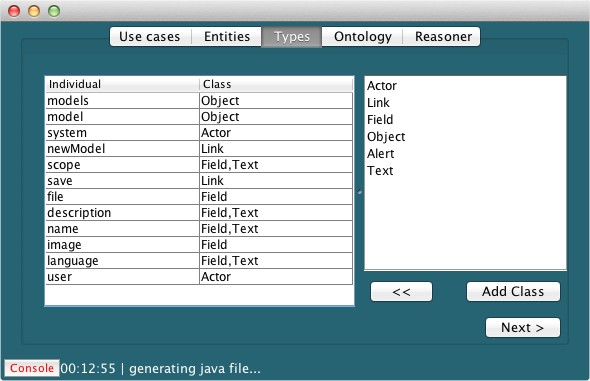}
		\caption{Classes to types association interface.}
		\label{figure:ss3}
	\end{figure}
	
	\begin{figure}[tb]\small
		\centering
		\includegraphics[width=45em]{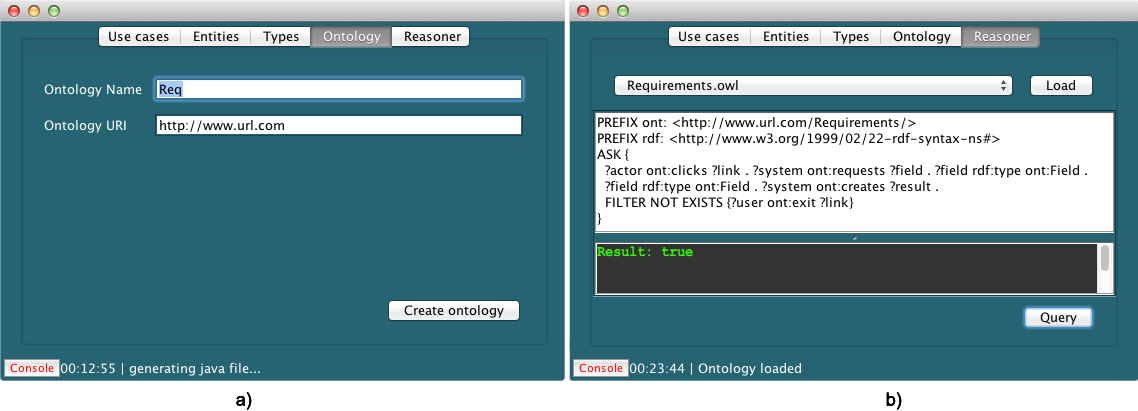}
		\caption{Create ontology and query interfaces.}
		\label{figure:ss4}
	\end{figure}
	\vspace{-0.5em}
	\subsection{Use case specification}
	\vspace{-0.3em}
	The tool supports the input of use case descriptions, in the textual format defined by the \ac{RUS} in use.
	In Figure~\ref{figure:ss1} \textbf{a)}, we present the tool's interface to specify the use case description.
	The user interface provides means to specify the user input and the system response.
	It is possible to see the model upload example (from Figure~\ref{figure:processed_usecase}) in our tool.
	In the left column, we specify the user input, and on the right, the system response.
	The user might select the validate button, in order to check the input against the \ac{RUS}. 
	When the generate button is pressed, the tool generates ANTLR (ANother Tool for Language Recognition) parsers, and extracts the \texttt{Individuals}, \texttt{Data properties} and \texttt{Object properties} from the requirements input. 
	It then proceeds to the next stage in the process (and the next screen: Entities).

	\subsection{Entities preview}
	\vspace{-0.3em}
	Once extracted, the inferred entities are shown to the user, in the interface which our tool provides to query them.
	In Figure~\ref{figure:ss1} \textbf{b)}, it is possible to see the extracted \texttt{Object properties} (relations) and the \texttt{Individuals}.
	Among others elements, we have the entities \texttt{user} and \texttt{newModel}, and the \texttt{clicks} relation, which were used to illustrate our approach.
	When the user finishes the tasks in this screen, it may select the \textbf{Types} tab or click the \textbf{Next} button, in order to proceed to the next screen. 	

	\subsection{Individuals' types}
	\vspace{-0.3em}
	The individuals' types information should be specified by the user, since we are not able to infer such information.
	Resorting to the interface depicted in Figure~\ref{figure:ss3}, the user should create the types (with the \textbf{add} button), and then assign one (or more) types to each individual.
	We have input a set of classes (\texttt{Actor}, \texttt{Link}, \texttt{Field}, \texttt{Object} and \texttt{Alert}), and associated them with the corresponding individuals.
	In this interface, we stated (for instance) that an \texttt{user} is an \texttt{Actor}.
	The classes are listed and created in the right side, and associated with the individuals listed on the left side.
	It is possible to associate several classes to the same individual, as is the case of the \texttt{name}, which is both an \texttt{Object} and a \texttt{Field}.
	\vspace{-0.5em}
	\subsection{Ontology generator and Reasoner}
	
	The last step performed in our tool, is to provide the missing informations to generate the ontology.
	The user must specify a valid url (in our case, \texttt{http://www.url.com}), and select the \textbf{Create Ontology} option.
	Such action generates the corresponding \ac{OWL} file, as presented in Figure~\ref{figure:ss4} \textbf{a)}.
	The file is a valid \ac{OWL} ontology, described in \ac{OWL} Manchester Syntax.
	We provide also a simple way to preview the generated file in its textual format. 
	That functionality is also available to preview the \ac{RUS} file.
	
	Furthermore, it is possible to query the generated ontology from within our tool.
	Figure~\ref{figure:ss4} \textbf{b)} depicts the interface which allows to write and execute \ac{SPARQL} queries.
	\vspace{-1.5em}
\section{Conclusions and Future Work}
\label{sect:c_fw}

	In this paper we have presented an approach to formalize use cases in OWL, in order to support the identification of requirements patterns in the use case descriptions. 
	Our approach started with the definition of a restricted natural language for requirements specification. 
	Such format, the \ac{RUS}, enables us to specify how user input should be expressed and how the resulting use cases can be transformed into an \ac{OWL} ontology. 
	The transformation rules provide us with most of the required information to create the ontology. 
	However, the process is not fully automated, as information such the types of specific individuals cannot typically be inferred. 
	We have implemented a prototype tool, which assists the user in inputing that missing information, at the same time as it shows the inferred knowledge. 
	Our tool is then able to create the \ac{OWL} ontology, with an associated instance. 
	Resorting to OWLs' query engines, we are able to perform queries over the ontology to identify requirements patterns in the use cases. 
	Our approach was illustrated with an example use case for uploading a model to a \ac{CMS}. 
	Resorting to this example we present the implemented tool, which supports our framework. 
	We also illustrated how requirements patterns could be identified in the example. 

	In order to further validate our approach we are interested in evaluating the expressiveness of the \ac{RUS}. 
	To do such, we will use the full requirements specification of existing software systems, and map them into our specification format. 
	This, besides enabling to assess the expressiveness of the language, will enables us to further develop the requirements patterns and their identification. 
	Indeed, our goal is to create a requirements patterns catalog by analyzing the requirements specifications of different software systems. 

	The proposed approach is part of a wider project which aims to create software prototypes, starting from use case specifications. 
	To help in this process, we propose to focus on a specific domain, resorting to the entities and workflow framework. 
	This feature is depicted in Figure~\ref{figure:ov_process}, in the arrows \texttt{e)} and \texttt{d)}. 
	By associating requirements patterns with architectural patterns, as presented in our framework in Figure~\ref{figure:ov_process} - transition \texttt{c)}, we propose to obtain hints about the final system implementation directly from the use cases. 
	The case studies above can also be of use in this. 
	By inferring architectural patterns presents in the use case specifications, we will be able to compare the actual implementations against the inferred patterns.

	Finally, while at the moment we are mainly interested in extracting the entities and their relationships from the use case specifications, it is also possible to extract other kinds of information from the use cases, specifically behavioral aspects. 
	We propose to analyze the possibility of extracting behavioral information from the use cases, for instance as \ac{CPN}, activity or state machine diagrams. 
	These models will allow us, not only to formally verify the use cases from a behavioral perspective, but also to simulate the use cases' flow, via animated diagrams. 
	This will help us to validate the use case specifications.	
\vspace{-1em}
\bibliographystyle{eptcs}
\bibliography{generic}
\end{document}